\documentclass{article}
\usepackage{epsfig}
\usepackage{amsmath}
\usepackage{amsfonts}\tolerance=10000
\date{\today} 
 
\usepackage{graphicx}

\newcommand{\insertplot}[5]{\begin{figure}
 \hfill\hbox to 0.05in{\vbox to #5in{\vfill
 \inputplot{#1}{#4}{#5}}\hfill}
 \hfill\vspace{-.1in}
 \caption{#2}\label{#3}
 \end{figure}}
 \newcommand{\inputplot}[3]{
 \special{ps: plotfile #1}
}
\newcounter{fig}

\newcommand{\beqe}{\begin{eqnarray}}
\newcommand{\eeqe}{\end{eqnarray}}

\numberwithin{equation}{section}
\newcommand{\be}{\begin{equation}}
\newcommand{\ee}{\end{equation}}
\newcommand{\bea}{\begin{eqnarray}}
\newcommand{\eea}{\end{eqnarray}}

\begin{document}

\title{Remarks on bell-shaped lumps: stability and fermionic modes} 
 
\author{{\large Y. Brihaye}$^{\dagger}$
and {\large T. Delsate}$^{\dagger }$ \\ \\
$^{\dagger}${\small Physique Th\'eorique et Math\'ematique, Universit\'e de
Mons-Hainaut, Mons, Belgium}\\
  }
\maketitle   
\begin{abstract}
We consider non-topological,  "bell-shaped" localized and regular  solutions available
in some 1+1 dimensional scalar field theories. Several properties of such solutions are studied,
namely their stability and the occurence of fermion bound states in the background of a kink and a kink-anti-kink
solutions of the sine-Gordon model.
\end{abstract}

\section{Introduction}
Non-linear phenomena play an important role in many sectors of physics. The most known classical solutions obeying
non-linear equations  are solitons and topological defects, see e.g. \cite{manton,vilenkin} for recent reviews. 
The stability of these objects is usually related to the non-trivial topology of the space of configurations
where the fields of the underlying physical models take their values.
Non-topological lumps are worth to be studied as well. Here topological 
arguments cannot be invoked to show stability but they can turn out to be stable through the occurence
of conserved bosonic currents related to global symmetries of the underlying model. 
On the other hand, some models can admit regular, localized, finite energy
classical but unstable solution which can be relevant in some specific contexts.
Sphalerons, which appear as unstable solutions in the electroweak model  \cite{sphaleron} are believed to play a role in the Baryon Asymmetry of the Universe \cite{rubakov}
and the computation of their normal modes \cite{brihaye_kunz1} is an important ingredient for the
evaluation of the rate of baryon/lepton-number violating processes.
Various extensions  of the sphaleron solution  have been emphasized. In a very recent 
paper, families of new axially symmetric solutions representing sphaleron-anti-sphaleron 
bound states have been constructed  \cite{kleihaus_kunz}.

Among all known non-linear lumps, the "kink-solution" available in the $1+1$-dimensional scalar
field theory with a $\lambda \phi^4$ potential is probably the simplest and the most popular.
Recently several authors have constructed different types of localized lump-solutions
in scalar field theories with specific families of the self-interaction potentials \cite{bazeia}. 
A special emphasis was set on "bell-shaped" lump characterized by $\lim_{x\to \pm \infty} \phi(x)=0$;
contrasting with the traditional kink-solutions.
%%%% which have $\lim_{x \to \infty}\phi(x) \neq \lim_{x \to -\infty}\phi(x)$.
In particular, the authors of \cite{bazeia} exhibit a family of potential admitting a lump
with the same profile as the celebrated Korteweg-de-Vries solution. A long list of
possible physical applications is given in \cite{bazeia}, among them, we point out the
possibility for these models to be applied in a brane-world involving a single extra dimension
and for the description of tachyonic states living on the brane, see e.g. \cite{brito} and references in this paper.
For such kind of applications, as well as for many other, the number of unstable modes
of the classical solution constitutes an essential ingredient. The question of stability, which to
our knowledge is not emphasized in
\cite{bazeia},  is considered in the second section of the present note.   The linear equation 
determining the normal modes turns out to be Quasi-Exactly-Solvable \cite{turbiner}. The normalizable solutions
of the equation can   be obtained  explicitely and one of the eigenvalues is negative, indicating
an instability.
The corresponding direction of instabity can be expressed in terms of elementary functions.

Recently again, kink and anti-kink solutions were used in a different physical context \cite{postma}.
In \cite{vachaspati}, the spectrum on the Dirac equation in the background of several solutions available
in the $\lambda \phi^4$ model  has been studied.
%%%to mimick  in a simpler model the problem of constructing fermionic
%%%bound states in the background of  
%%%torus-like configurations of cosmic strings (a so called vorton\cite{lemperiere}). 
The investigation of fermionic bound states in the background of classical bosonic fields
(solitons or sphalerons) has a long history.
 The Dirac equation in the background of 
a  linear cosmic string  was studied  in \cite{rossi}, \cite{weinberg}.  The existence of normalisable fermion zero-mode
  in the background of a sphaleron was adressed in \cite{nohl},\cite{boguta}. 
   Further studies of fermionic
  bound states and level crossing  were performed in \cite{ringwald}, the spectrum along  the full non contractible loop passing through the sphaleron 
  was reported in \cite{brihaye_kunz2}.

In the third section of this paper, we reconsider the calculation of \cite{vachaspati} by replacing the kink 
and kink-anti-kink  backgrounds by their analogues available in the sine-Gordon model.
The advantage of the sine-Gordon trigonometric potential over the quartic polynomial one is that 
sine-Gordon model admits an explicit solution describing  fully the kink-anti-kink interaction; 
as a consequence, the spectrum of the Dirac equation in this background
can be computed over the full range of configurations, 
including the region of the parameters where kink and  the anti-kink interacts with each other. 
The qualitative properties of the analysis reported in \cite{vachaspati} are confirmed with the sine-Gordon model
for well separated lumps.
However  the solutions of the sine-Gordon equation allow for  configurations where the  two lumps are 
close to each other.  The spectral analysis can be extended to this case as well. 
Our results demonstrate how fermions bound states
can emerge from the continuum when the wall and the anti-wall become  well separated or, equivalently,
how existing fermion bound states are absorbed into the continuum when the two lumps come close to each other.
Some conclusions and perspective are mentionned in Sect.4. For completeness,
the definition of the Poschl-Teller equation
is reminded in an Appendix.
%%%%%%%%%%%%%%%%%%%%%%%%%%%%%%%%%%%%%%%%%%%%%%%%%%%%%%%%%%%%%%%%%%
\section{Bell-shaped lumps: Stability}
In this section we study the stability of several new lump solutions
which were presented recently in \cite{bazeia}. The models considered are
1+1 dimensional scalar field theory of the form
\be
    {\cal L} = \frac{1}{2} \partial_{\mu}\phi \partial^{\mu}\phi - V(\phi) 
\ee  
with suitable potential $V(\phi)$. 
For static fields, the
classical equation has the form $d_x^2 \phi = \frac{d V}{d \phi}$.
The linear stability  of a 
 solution, say $\phi_c(x)$,  of  this equation,  can be studied by diagonalizing the quadratic 
 part of the lagrangian in a perturbation around $\phi_c$. 
 Concretely, one has to compute the eigenvalues allowing for
 normalisable eigenfunctions of the spectal equation
\be
     ( -\frac{d^2}{dx^2} + \left. \frac{d^2 V}{d \phi^2} \right|_{\phi=\phi_c})\eta = \omega^2 \eta \ \ \ , \ \ \ 
     \phi(t,x) = \phi_c(x) + e^{i \omega t} \eta(x)
\ee
Negative eigenvectors $\omega^2 < 0$, reveal the existence of instabilities of the classical solution $\phi_c$.
\\
\\
{\bf Case 1: $\phi^4$-Potential}
\\
\\
The case of a quartic potential is well known but we mention it for completeness
and for the purpose of comparaison of the eigenmodes with the cases mentionned next.
The potential and the corresponding kink-solution read
\be
\label{kink}
    V(\phi) = \frac{1}{2} (\phi^2 - 1)^2 \ \ \ , \ \ \ \phi_c(x) = \pm \tanh(x)
\ee
and the stability equation corresponds to the Poschl-Teller equation (see Appendix) with $N=2$ and
$\omega^2 = \omega_p^2 +4$. The eigenvalues are then  $\omega^2= 0,3,4$. It is well know that
the $\lambda \phi^4$-kink  has no negative mode and is stable. The corresponding eigenfunctions are
%%%%modes corresponds to
\be
    \eta_0 = \frac{1}{\cosh(x)^2} = - \frac{d \phi_c}{dx} \ \ , \ \
    \eta_3 = \frac{\sinh(x)}{\cosh^2(x)} \ \ , \ \ 
    \eta_4 = \tanh^2(x) - \frac{1}{3} \ \ . 
\ee
The zero mode corresponds to infinitesimal translations of the classical solution $\phi_c$ in the space variable.
\\
\\
{\bf Case 2: Inverted $\phi^4$-Potential}
\\
\\
This case  corresponds to a quartic potential  with inverted sign.
The potential and the corresponding lump are given by
\be
    V(\phi) = \frac{1}{2} \phi^2 (1- \phi^2) \ \ \ , \ \ \ \phi_c(x) = \pm \frac{1}{\cosh(x)} \ .
\ee
In this case also, the stability equation corresponds to the Poschl-Teller equation with $N=2$ but with and
different shift of the eigenvalue, for instance
$\omega^2 = \omega_p^2 +1$; leading to $\omega^2= -3,0,1$. As a consequence
the lump is characterized by one negative mode and one zero mode. Up to a normalisation
they are given by
\be
%%%    
      \ \ \tilde\eta_{-3} = \frac{1}{\cosh(x)^2} \ \ \ , \ \ \ 
      \ \ \tilde \eta_0 = \frac{\sinh(x)}{\cosh^2(x)} = -\frac{d \phi_c}{dx}
      \ \ , \ \ \tilde \eta_1 = \tanh^2(x) - \frac{1}{3}
\ee
as usual, the zero mode is associated with the translation invariance of the underlying field theory.
\\
\\
{\bf Case 3: Cubic $\phi^3$-Potential}
\\
\\
The last case investigated corresponds to a  potential of third power in the field $\phi$.
The potential and the corresponding lump are given by
\be
    V(\phi) = 2 \phi^2 (1- \phi) \ \ \ , \ \ \ \phi_c(x) = \pm \frac{1}{\cosh^2(x)} \ .
\ee
The correspondind  normal mode equation corresponds to the Poschl-Teller equation with $N=3$ with an
appropriate shift of the spectrum~:
$\omega^2 = \omega_p^2 +4$; leading to $\omega^2=-5,0,3,4$. 
The lump here is then characterized by one negative mode, one zero mode and two positive modes. 
Up to a normalisation
they are given by
\be
%%%     
      \ \ \tilde\eta_{-5} = \frac{1}{\cosh^3(x)} \  , \  
     \eta_0 = \frac{\sinh(x)}{\cosh^3(x)} = -\frac{1}{2}\frac{d \phi_c}{dx} \ , \ 
     \eta_3 = \frac{1}{\cosh^2(x)}( \tanh^2(x) - \frac{1}{5}) \  , \  
     \eta_4 = \tanh(x) ( \tanh^2(x) - \frac{3}{5})
\ee
In \cite{bazeia},  deformations  of the three potentials given above are studied as well
(e.g. is cases 1 and 3, deformations breaking the reflection symmetry
$\phi \to -\phi$). The corresponding stability equation
is not of the Poschl-Teller type but their spectrum could  be studied perturbatively from the ones obtained above. 
It should be interesting
to see in particular  how the unstable mode (or the zero mode in the case 1) 
evolve in terms of the  coupling constant parametrizing the deformation.   
\\
%%%%%%%%%%%%%%%%%%%%%%%%%%%%%%%%%%%%%%%%%%%%%%%%%%%%%%%%%%%%%%%%%
%%%%%\section{Stability}
%%%%%%%%%%%%%%%%%%%%%%%%%%%%%%%%%%%%%%%%%%%%%%%%%%%%%%%%%%%%%%%%%%
%%%%%%%%%%%%%%%%%%%%%%%%%%%%%%%%%%%%%%%%%%%%%%%%%%%%%%%%%%%%%%%%%%
\section{Fermion modes in a Sine-Gordon Kink-anti-Kink system}
%%%%%%%%%%%%%%%%%%%%%%%%%%%%%%%%%%%%%%%%%%%%%%%%%%%%%%%%%%%%%%%%%%
Another aspect of lump-like structures of the type discussed here is the possibility
to couple them to a fermion field and to generate a mass to the fermion through a Yukawa interaction
of the fermion to the scalar field $\phi$. Using the notations of Chu-Vachaspati \cite{vachaspati}, we have
\be
\label{lagrangian_full}
 {\cal L} = \frac{1}{2} \partial_{\mu}\phi \partial^{\mu}\phi - V(\phi) 
 + i \overline \psi \gamma^{\mu} \partial_{\mu} \psi 
   - g (\phi-C) \overline \psi \psi 
\ee
where $\psi$ is a two-component spinor and the two-dimentional Dirac matrices can be chosen in terms
the Pauli matrices $\gamma^0 = \sigma_3$, $\gamma^2=i \sigma_1$, $g$ is the Yukawa coupling constant
and the shift constant $C$ has to be chosen appropriately 
($C=0$ for the $\phi^4$ model, $C = \pi$ for the sine-Gordon model).
To solve the Dirac equation in the background of a classical solution $\phi_c$ 
available in the bosonic sector of the model, it is convenient to parametrize \cite{vachaspati}
the Dirac spinor according to 
\be
   \psi = \left( \psi_1  , \psi_2 \right)^t  \ \ , \ \   \psi_1 = e^{-iEt} (\beta_+ - \beta_-) \ \ , \ \ \psi_2 = e^{-iEt} (\beta_+ + \beta_-) \  \ \
     .
\ee
where $\beta_{\pm}$ are functions of $x$. 
Then the system of first order Dirac equations is transformed into two decoupled second order
equations for $\beta_-$ and $\beta_+$~:
\be
\label{dirac}
       (-\partial_x^2 + V_{\pm}(\phi_c)) \beta_{\pm} = E^2 \beta_{\pm} \ \ \ , \ \ \ 
       V_{\pm} (\phi_c) = g^2 \phi_c^2 \mp g \partial_x \phi_c
\ee
The Dirac equation in the background if the one Kink corresponding to the model (\ref{kink}) was studied in 
detail in \cite{vachaspati}, it was found in particular that the equations (\ref{dirac}) are Poschl-Teller
equations respectively with $N=g$ and $N=g-1$ and $ \omega_p^2 = E^2- g^2$. 
For any value of $g$ such that $n-1 < g \leq n$
there exist $n$ fermionic modes given by
\be
           E_j = \sqrt{j(2g-j)} \ \ , \ \ j = 0,1, \dots , n-1
\ee
At each integer value of $g$, a supplementary bound state emerge from the continuum $E=g$ and exist for 
$g > n$.

Next, the Dirac equation was studied in the background of a Kink-anti-Kink configuration 
say $\phi_{K\overline K}$
represented by mean of superposition of a Kink centered at $x = -L$ and an anti-Kink centered at $x=L$ with $L >> 1$~:
\be
\label{superposition}
      \phi_{K\overline K}(x) = \tanh(x+L) - \tanh(x-L) - 1 \ \ . 
\ee
Such configurations are also "bell-shaped". 
The main result is that the spectrum of the Dirac equation in the background of $\phi_{K\overline K}$
deviates only a little from the spectrum available in the background of a single Kink (or an anti-Kink) for
$L \gg 1$. The analysis of the spectrum in the region $L\sim 1$ is unreliable since the linear superposition
(\ref{superposition}) is an approximate solution of the classical equation only for large values of $L$.
\begin{figure}
\centering
\epsfysize=8cm
\mbox{\epsffile{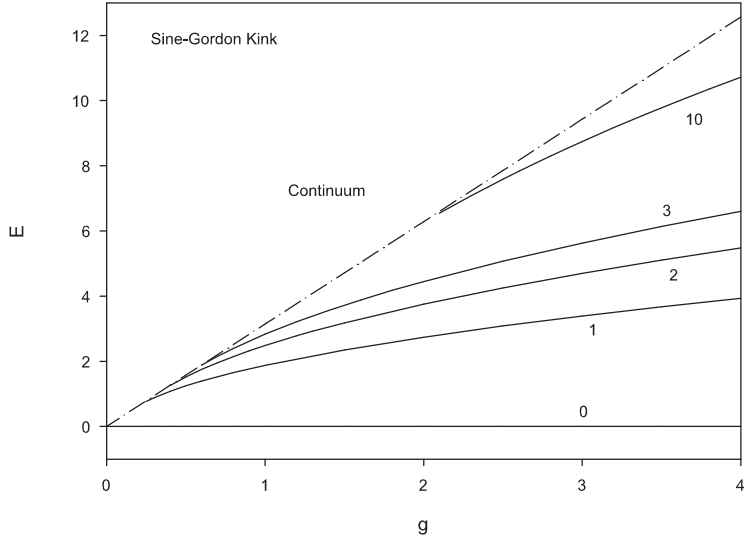}}
\caption{\label{fig1}
The fermionic bound state corresponding to the single sine-Gordon kink as function
of the constant $g$. The level $n=0,1,2,3$ and $10$ are represented}
\end{figure}  
%%%%
\begin{figure}
\centering
\epsfysize=8cm
%%%%\mbox{\epsffile{fig2.eps}}
\mbox{\epsffile{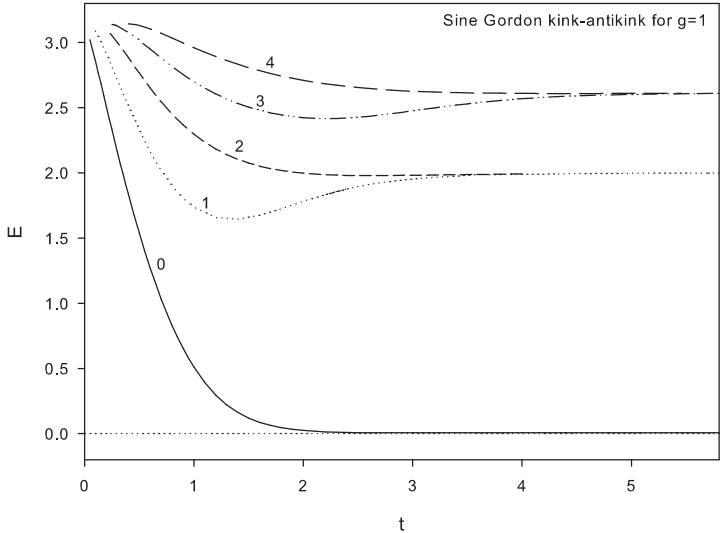}}
\caption{\label{fig2}
The fermionic bound state corresponding to the sine Gordon kink-anti-kink solution
 as function of time for $g=1$ and $u=0.5$. The level $n=0,1,2,3,4$ are represented}
\end{figure}
%%%%%%%  
In this paper we have reconsidered the Dirac equation (\ref{dirac}) for the Kink and Kink-anti-Kink 
solutions available
in the Sine Gordon model, the potential and the form of the fundamental lump in this model are
\be
V(\phi) = \frac{(1- \cos(\phi))}{2} \ \ \ , \ \ \ \phi(x) = \pm 4 {\rm Atan}(e^x) \ .
\ee
On Fig. 1, a few fermionic bound states are represented as function of the parameter $g$
(for clarity, we omitted the modes $n=4,\dots 9$ on the graphic). The figure
present exactly the same pattern than in the case of the $\phi^4$-kink. In particular, new bound states
emerge regularly from the continuum at critical values of $g$, say $g=g_n$. 
Let us mention 
that the normal modes about the sphaleron and bisphaleron solutions \cite{brihaye_kunz1} of the
standard model of electroweak interactions also lead to a similar pattern. 

Because these qualitative properties of the solutions are similar to the one of the $\phi^4$-kink we expect
the features discovered in \cite{vachaspati} to hold in the case of well separated sine-Gordon
kink and anti-kink. In the sine-Gordon model, we can take advantage of the fact that
 an exact form of the kink-anti-kink solution
is available~:
\be
\label{kak}
     \phi_{K \overline K} = 4 {\rm Atan} \frac{\sinh(ut/\beta)}{u\cosh ( x/\beta)} \ \ , \ \ \beta = \sqrt{1-u^2}
\ee
where $u$ is a constant related to the relative velocity of the two lumps. 
We used this solution to study the Dirac equation in the background of a moving wall-anti-wall
system (approaching or spreading each other, according to the sign of $t$) and were able to
study the spectrum of the bound states in the domain of the parameter
$t$ become small, i.e. when the kink and the anti-kink interact. Solving Eq. (\ref{dirac}) in the background
(\ref{kak}), we implicitely assume that the motion of the kink and of
the anti-kink is treated adiabatically. A time-integration of the full equations will be reported elsewhere
\cite{delsate}. 
\begin{figure}
\centering
\epsfysize=8cm
%%%%\mbox{\epsffile{fig2.eps}}
\mbox{\epsffile{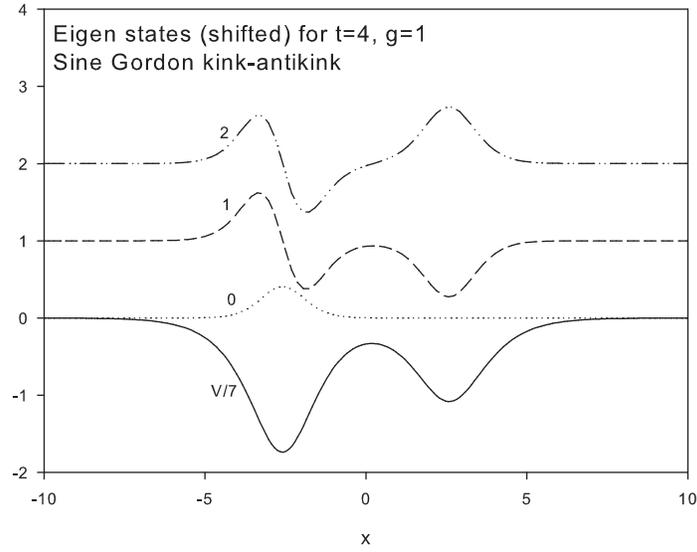}}
\caption{\label{fig3}
The profile of the potential and of the first three fermionic bound states 
corresponding to the single sine Gordon kink-anti-kink solution
for $t=4$, $g=1$ and $u=0.5$ are represented}
\end{figure}  
%%%%%%%%%%%%%%%%%%%%%%%%%%%%%%
\begin{figure}
\centering
\epsfysize=8cm
%%%%\mbox{\epsffile{fig2.eps}}
\mbox{\epsffile{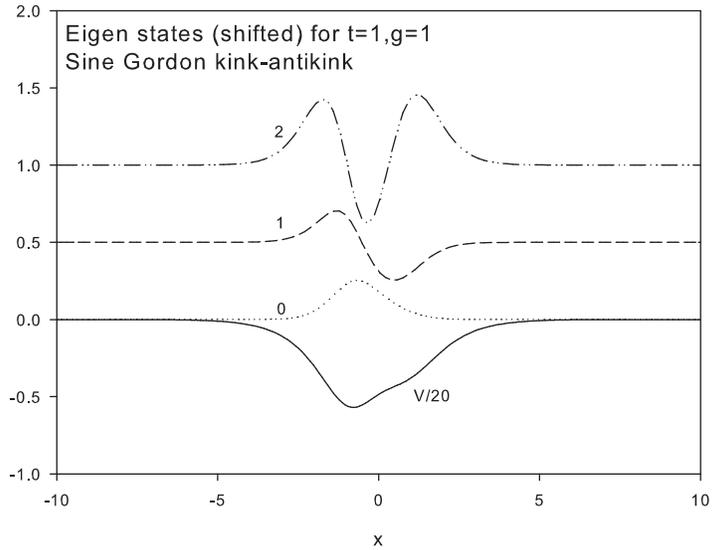}}
\caption{\label{fig4}
The profile of the potential and of the first three fermionic bound states 
corresponding to the single sine Gordon kink-anti-kink solution
for $t=1$, $g=1$ and $u=0.5$ are represented}
\end{figure}  
The results are summarized on Fig. 2 where the fermionic eigenvalue $E$ for the ground state $n=0$ and the first few excited states are represented as function of $t$.
Our numerical integration of the spectral equations (\ref{dirac}) for several values of $t$
reveals that only the ground state subsists
in the $t\to 0$ limit where it enters in the continuum (this could be expected since the classical
background solution vanishes in this limit) and that the excited states join the continuum at finite
 values of $t$, depending  on the level of the exitation.
 It should be pointed out that the fermionic eigenvalue of the fundamental solution (line $n=0$)
 approaches $E=0$
 for $t \to \infty$ although staying positive. It decays roughly like $E \sim \exp -4t$. 
  The fact that the excited energy levels become pairwise degenerate in the large $t$-limit can be understood from
 the form of the potential. Indeed, for $t \gg 1$, the potential posseses two well separate valleys
 centered about the points $x \sim \pm ut$. In the neighbourhood of $x=\pm ut$, the form
 of the potential is given by $V_{\pm} = g^2 \phi^2 \mp g \phi'$. 
 Far away from the two regions of the $x$-line situated around $x=\pm ut$ (i.e in the region of the origin and in the
 asymptotic regions), the potential  is exponentially  small.
 In fact the effective potential under investigation results from a superposition
 of two (suitably shifted) potentials   which are supersymmetric
 partners of each other;  accordingly they have the same spectrum apart from the ground state of $V_+$.
 It is one of the striking property of supersymmetric quantum mechanics (see \cite{khare} for a review)
  that if $V_{\pm}(x)$ are supersymmetric partner potentials,
 the $k$-th energy level of $V_-$ coincides with the $k+1$-energy level of $V_+$. As
 a consequence, in our case, two 
 eigenvectors exists with roughly the same eigenvalue when the wall and the anti-wall are well separated, like e.g. in Fig. 3.
  While $t$ decreases, the two valleys of the potential
 have tendency
 to merge in the region of the origin, see Fig. 4, and the degeneracy of the two energy levels
 corresponding to each other by the supersymmetry is lifted, as shown on Fig. 2.
 It is tempting to say that the supersymmery of the spectum occuring for $|t| \gg 1$ is broken
 in the $|t| \to 0$-limit.
Said in other words, it turns out that when the wall and anti-wall approach each other,
the system cannot support fermion bound states and their spectrum merge into the continuum of the Dirac
equation. Seen opposite, fermion bound states (ground state and a number of excited modes, the number of them
depending on the coupling constant $g$) can emerge from the continuum when the wall and anti-wall separate
from each other.  We plan to study this phenomenon in more details by solving the full 
 Dirac-sine-Gordon
equation \cite{delsate}

 \section{Conclusion and perspectives}
 The stability equations associated with the bell-shaped lumps recently discussed in \cite{bazeia}
 turn out to belong to the class of quasi-exactly-solvable equation \cite{turbiner}. In particular
 their unstable mode can be computed explicitely. Accordingly, the underlying field theory can be used,
 along with \cite{brito}, as toy models for studying tachyonic modes in brane world.
 In Sect. 3, the study of the spectrum of the Dirac equation in the background of wall-anti-wall
 configuration has revealed the existence  of several number of fermionic bound states  (the exact
 number depending on the coupling constant) emerging from the continuum while the two walls get
 more separated is space. 
 We have pointed out the close relation of the pattern of eigenmodes with supersymmetric quantum mechanics.
 It would be tempting to use this feature in a context of higher dimensional space-time
 where the the extra-dimension would be the interpreted as the "spatial" coordinate of the walls. 
 Alternative mechanisms of fermion-brane interaction could be looked for in this direction. The main result in this topic
 was obtained in the celebrated achievement of \cite{rubakov}, 
 here a kink is used to localize the fermion in a d=5 space-time. 
 Generalisations of this idea to space-times inolving more than two co-dimensions have been emphasized,
  namely in \cite{nahm}, where topological solitons
 available in Yang-Mills and sigma-models are used as localizing mechanisms for fermions. 

Finally, let us point out that a lot of activity is devoted actually to the study of the interactions of solitons,
or of more general spatially localized objects, with themselves 
\cite{forgacs,vachaspati2}. 
The interaction of kink-solution with an external field of
radiations is studied in detail in \cite{forgacs}.  It is shown that the radiation exerts a negative pressure
on the kink and that, accordingly, the kink is pushed backwards. A similar analysis could be performed
in the case of wall-anti-wall configurations like the ones emphasized in \cite{vachaspati} or in the present paper.

%%%%%%%%%%%%%%%%%%%%%%%%%%%%%%%%%%%%%%%%%%%%%%%%%%%%%%%%%%%%%%%%%%
\section{Appendix: The Poschl-Teller Equation} 
The Poschl-Teller equation is known as the following one-dimensional
eigenvalue Schrodinger  equation for the corresponding Poschl-Tellel potential. 
\be
  -\frac{d^2}{dx^2} \eta - \frac{N(N+1)}{\cosh^2 x} \eta = \omega_p^2 \eta
\ee 
It is considered
on an appropriate domain of the Hilbert space of square integrable functions on the real line.
It is a standard result that, for integer values of $N$, the above equation
admits $N+1$ eigenvalues and eigenvectors which can be computed algebraically.
For the first few values of $N$ the eigenvalues are given by
\be
     \ \ \ N=1 : \omega_p^2 = -1,0 \ \ \ , \ \ \  
    N=2 : \omega_p^2 = -4,-1,0 \ \ \ , \ \ \ N=3 : \omega_p^2 = -9,-4,-1,0
\ee  
\\
\\
 \newpage
 \begin{small}
 
%%%%%%%%%%%%%%%%%%%%%%%%%%%%%%%%%%%%%%%%%%%%%%%%%%%%%%%%%%%%%%%%%%%%%%%%%%%%%%
 \end{small}

\end{document}